\documentclass[aps,pre,superscriptaddress,twocolumn,10pt]{revtex4-2}

\usepackage{xcolor,graphicx}
\usepackage{amssymb,amsmath,bbm}
\usepackage[utf8]{inputenc}
\usepackage{comment}
\usepackage{braket}
\usepackage[normalem]{ulem}
\usepackage{hyperref}
\hypersetup{colorlinks=true,linkcolor=blue,citecolor=blue,urlcolor=blue}

\begin{document}

\title{Energy exchange statistics and fluctuation theorem for non-thermal asymptotic states}

\author{Santiago Hern\'andez-G\'omez}
\thanks{These authors contributed equally to this work.}
\affiliation{CNR-INO, via Nello Carrara 1, I-50019 Sesto Fiorentino, Italy}
\affiliation{Research Laboratory of Electronics, Massachusetts Institute of Technology, Cambridge, MA 02139, USA}
\affiliation{European Laboratory for Non-linear Spectroscopy (LENS), Universit\`a di Firenze, I-50019 Sesto Fiorentino, Italy}

\author{Francesco Poggiali}
\thanks{These authors contributed equally to this work.}
\affiliation{CNR-INO, via Nello Carrara 1, I-50019 Sesto Fiorentino, Italy}
\affiliation{European Laboratory for Non-linear Spectroscopy (LENS), Universit\`a di Firenze, I-50019 Sesto Fiorentino, Italy}

\author{Paola Cappellaro}
\affiliation{Research Laboratory of Electronics, Massachusetts Institute of Technology, Cambridge, MA 02139, USA}
\affiliation{Department of Nuclear Science and Engineering, Department of Physics, Massachusetts Institute of Technology, Cambridge, MA 02139, USA}

\author{Francesco S. Cataliotti}
\affiliation{European Laboratory for Non-linear Spectroscopy (LENS), Universit\`a di Firenze, I-50019 Sesto Fiorentino, Italy}
\affiliation{CNR-INO, Largo Enrico Fermi 6, I-50125 Firenze, Italy}
\affiliation{Dipartimento di Fisica e Astronomia, Universit\`a di Firenze, via Sansone 1, I-50019 Sesto Fiorentino, Italy}

\author{Andrea Trombettoni}
\affiliation{Dipartimento di Fisica, Universit\`a di Trieste, Strada Costiera 11, I-34151 Trieste, Italy}
\affiliation{SISSA, via Bonomea 265, I-34136 Trieste, Italy}
\affiliation{INFN, Sezione di Trieste, via Valerio 2, I-34127 Trieste \& CNR-IOM DEMOCRITOS Simulation Center, via Bonomea 265, I-34136 Trieste, Italy}

\author{Nicole Fabbri}
\affiliation{CNR-INO, via Nello Carrara 1, I-50019 Sesto Fiorentino, Italy}
\affiliation{European Laboratory for Non-linear Spectroscopy (LENS), Universit\`a di Firenze, I-50019 Sesto Fiorentino, Italy}

\author{Stefano Gherardini}
\affiliation{European Laboratory for Non-linear Spectroscopy (LENS), Universit\`a di Firenze, I-50019 Sesto Fiorentino, Italy}
\affiliation{CNR-INO, Largo Enrico Fermi 6, I-50125 Firenze, Italy}
\affiliation{SISSA, via Bonomea 265, I-34136 Trieste, Italy}

\begin{abstract}
Exchange energy statistics between two bodies at different thermal equilibrium obey the Jarzynski-W\'ojcik fluctuation theorem. The corresponding energy scale factor is the difference of the inverse temperatures associated to the bodies at equilibrium. In this work, we consider a dissipative quantum dynamics leading the quantum system towards a, possibly non-thermal, asymptotic state. To generalize the Jarzynski-W\'ojcik theorem to non-thermal states, we identify a sufficient condition ${\cal I}$ for the existence of an energy scale factor $\eta^{*}$ that is unique, finite and time-independent, such that the characteristic function of the exchange energy distribution becomes identically equal to $1$ for any time. This $\eta^*$ plays the role of the difference of inverse temperatures. We discuss the physical interpretation of the condition ${\cal I}$, showing that it amounts to an almost complete memory loss of the initial state. The robustness of our results against quantifiable deviations from the validity of ${\cal I}$ is evaluated by experimental studies on a single nitrogen-vacancy center subjected to a sequence of laser pulses and dissipation. 
\end{abstract}

\maketitle

\section{Introduction}

The probability distribution of a random variable defined at two times generally depends on the details of the system's dynamical process. A key example is provided by the the internal energy variation $\Delta E(t)$. According to the formalism of stochastic thermodynamics~\cite{Hekking13,Campisi15,Elouard17,GherardiniQST2018,ManzanoPRX2018,GherardiniReview2022}, in a single quantum trajectory the value of $\Delta E(t)$ in the time interval $[t_0,t]$ is defined as the difference of eigenvalues of the quantum system Hamiltonian $\mathcal{H}(t)$ evaluated at the times $t_0$ and $t$. The fluctuations can be analyzed by means of the probability distribution ${\rm P}(\Delta E(t))$, or its characteristic function $\mathcal{G}(u,t)$, with $u$ a complex number, given by the Fourier transform of ${\rm P}(\Delta E(t))$.

Fluctuation theorems can connect non-equilibrium quantities to equilibrium properties of the dynamical system~\cite{Esposito09,Campisi11}. In the following we are going to address \emph{energy exchange} fluctuation theorems for dissipative quantum systems. In the original formulation of exchange fluctuation theorems~\cite{Jarzynski04}, two bodies $\mathcal{A}$ and $\mathcal{B}$ with finite dimension are prepared in two equilibrium states at temperatures $T_\mathcal{A} \equiv 1/(k_{B}\beta_{\mathcal{A}})$ and $T_\mathcal{B} \equiv 1/(k_{B}\beta_{\mathcal{B}})$ respectively, and then placed in thermal contact for a given time interval. 
Fluctuation theorems address the question whether there exist a (single) value of $u=j\eta$ such that the characteristic function (for the separated system) becomes independent of time and equal to 1. Then, such a value acts as an effecting `macroscopic' rescaling at all times of the heat exchange fluctuations between the two bodies.
Once they are divided again, the question was to understand whether there exist a single `macroscopic' quantity able to re-scale at any time the statistics (thus, the fluctuations) of the heat exchange between the two bodies. In \cite{Jarzynski04} under the assumption of weak interaction between the two bodies, it was shown that for any time after physical operation of decoupling the systems it is
\begin{equation}\label{eq:JWE}
\langle e^{-\Delta\beta \, Q}\rangle = 1 \,,  
\end{equation}
where $Q$ is the stochastic value of the heat exchanged by the two bodies, and the average $\langle\cdot\rangle$ is performed over the exchanged heat distribution. In Eq.~(\ref{eq:JWE}), $\Delta\beta \equiv \beta_{\mathcal{B}} - \beta_{\mathcal{A}}$ denotes the energy scale factor that normalizes the heat fluctuations. The exchanged heat equality discussed above is valid~\cite{Hovhannisyan_NoGo_21} also by taking an initial thermal state and employing the so-called two-point measurement (TPM) scheme~\cite{Tasaki00,Talkner07}, where quantum measurements are performed at the initial and final time. The TPM scheme is a convenient way to access fluctuations, as experimentally shown in Refs.~\cite{Smith18,Pal19,HernandezGomez20, HernandezGomez21,Cimini20,RibeiroPRA2020,Solfanelli2021}. 
Recent activity focused on how energy exchange statistics could be reconstructed using schemes beyond the TPM~\cite{Micadei20,Gherardini20x,SonePRL2020,MicadeiPRL2021,hernandez2022experimental,HernandezGomez23,Maeda23}, when quantum coherences in the initial state are taken into account.

Noticeably, Eq.~(\ref{eq:JWE}) is valid also when a (finite-dimension) body ${\cal A}$, initially in the equilibrium state at temperature $T_{\cal A}$, is put at a certain time $t^*$ in contact with a \textit{thermal bath} ${\cal B}$ at temperature $T_{\cal B}$~\cite{Ramezani18}. The detailed balance condition (DBC) is a sufficient condition for the validity of a Crook fluctuation relation~\cite{Crooks99}; the converse, instead, is not necessarily valid~\cite{Ramezani18}. Independently of the microscopic details of the coupling between ${\cal A}$ and ${\cal B}$, Eq.~(\ref{eq:JWE}) is valid at any time later than $t^*$~\cite{Ramezani18}. 
This shows that $\Delta \beta$ plays the role of a macroscopic quantity that fully characterizes fluctuations of the energy exchanged at any time, and it is solely determined by the knowledge of the initial and final equilibrium states of the thermodynamic process. 
In general, this symmetry in the exchanged heat statistics is not present. As an example, in \cite{JevticPREexchange}, Eq.~(\ref{eq:JWE}) is generalized to the case the initial state of the two bodies $\mathcal{A}$ and $\mathcal{B}$ are non-thermal as an effect of classical correlations. However, the derivation of a fluctuation relation as Eq.~(\ref{eq:JWE}) in closed-form for such a case requires the complete knowledge of both bodies and the full access to their whole dynamics. Whether and when one can determine a counterpart of $\Delta\beta$ for non-thermal initial and final states is an open question.

In this paper, we present a fluctuation theorem for the energy exchanged by a quantum system with a dissipative environment, not necessarily at thermal equilibrium. We introduce a sufficient condition for the existence of a unique, finite and constant (time-independent) energy scale factor $\eta^{*}$ such that the characteristic function $\mathcal{G}(u,t) \equiv \int {\rm P}(\Delta E(t))e^{ju\Delta E(t)}d\Delta E$ obeys the relation 
\begin{equation}\label{eq:QR_steady_states} 
\mathcal{G}(j\eta^{*},t) = \langle e^{-\eta^{*}\Delta E(t)}\rangle = 1 \quad \text{for} \quad t \geq t^{*} \,.
\end{equation}
In Eq.~(\ref{eq:QR_steady_states}), $j$ denotes the imaginary unit. In general, if the energy scale factor is chosen without a specific criterion, then $\mathcal{G}(j\eta,t)$ [with $\eta\in\mathbb{R}$] is not equal to $1$ and its value depends on the choice of both the initial state and the whole dynamical process for any time $t^*$.

The energy exchange fluctuation theorem (\ref{eq:QR_steady_states}) for non-thermal asymptotic states and the discussion of the sufficient condition making it valid are the main results of the paper. 
This sufficient condition contains the DBC under two working hypotheses, namely, evaluating fluctuations by means of the TPM scheme, thanks to which a classical-quantum correspondence is possible~\cite{JarzynskiPRX2015}, and a dynamical process that asymptotically leads the system into an asymptotic state. 
A key point of the sufficient condition is related to the memory loss of the exchanged energy from the initial quantum state. 
For this reason, we motivate the statement of the sufficient condition by making use of Stern-Gerlach protocols~\cite{Gerlach1922,Castelvecchi2022,Mello2014}, where the memory loss is induced by quantum measurements. In this way, we can better illustrate the assumptions behind the fluctuation theorem and its physical interpretation.

Eq.~(\ref{eq:QR_steady_states}) can find application also to quantum systems under dissipation that do not reach an equilibrium but admit a non-equilibrium steady-state, thus extending the findings of Ref.~\cite{HatanoPRL2001} to the case of dissipative maps. As an example, one may consider the case of time evolutions along isoenergetic trajectories, whereby asymptotically the probabilities to measure the energy values of the system neither depend on time nor on the initial state.

Finally, the robustness of the fluctuation theorem \eqref{eq:QR_steady_states} is tested under exemplary experimental conditions where the validity (or not) of the fluctuation theorem's assumption, as well as of the DBC, can be controlled. We also discuss how $\eta^{*}$ depends on the values taken by the parameters of the Hamiltonian. Here, the used platform is an autonomous dissipative Maxwell demon formed by a Nitrogen-Vacancy (NV) center in diamond at room temperature, introduced in~\cite{HernandezGomezPRXQuantum22}.

\section{Energy scale factor in the asymptotic regime}
\label{app:varepsilon_formalism}

Let us discuss the working hypotheses that will be used in the rest of the paper. 

We consider a quantum system with finite dimensional Hilbert space (having dimension $n$) subjected to a Hamiltonian with time-independent eigenvalues, and at the same time under the effect of dissipation. We assume that the dynamics is determined by a completely positive trace-preserving (CPTP) map $\Phi$ that for large times leads the system to an asymptotic, possibly non-equilibrium, state of the form 
\begin{equation}\label{eq:asymptotic_nonequilibrium_state}
\lim_{t\to\infty} \Phi_t[\rho_0] = \sum_{f=1}^n P_{f}(\infty) \Pi_{f}(\infty) + \chi[\rho_0,t] \, ,
\end{equation}
with fixed values on the diagonal with respect to the energy basis $\{E_{f}\}$, with $\Pi_{f} \equiv |E_{f}\rangle\!\langle E_{f}|$, and possibly time-dependent off-diagonal terms $\chi$. The probabilities $P_{f}(\infty)$, for $f \in \{1,\dots,n\}$, are independent of both the time and the generic initial state $\rho_0$, while $\chi$ can depend on $\rho_0$ even in the asymptotic regime.

Exchange energy fluctuations are accessed by employing the TPM scheme. We recall that the TPM scheme provides complete energy exchange statistics in terms of correct marginals and linearity in the state, for any initial state that is diagonal in the system Hamiltonian~\cite{LostaglioKirkwood2022}.

In order to demonstrate the existence of a unique, finite, time-independent energy scale factor $\eta^*$ obeying the fluctuation theorem (\ref{eq:QR_steady_states}), as first step we prove the following Lemma that is valid in the asymptotic limit:

\vspace*{0.15cm}
\noindent
{\bf Lemma}~[\emph{Uniqueness of a non-trivial energy scale factor in the asymptotic limit}]: For a quantum system with a Hamiltonian with time-independent eigenvalues and subjected to a dissipative map, the equality $\mathcal{G}(j\eta,t)=1$ is fulfilled asymptotically for $t \rightarrow \infty$ only for two distinct values of $\eta$ at most, i.e., $\eta=0$ (trivial solution) and $\eta=\eta^{*}$, with $\eta^{*}\in\mathbb{R}$ a finite real number depending only on the initial and the asymptotic states.
\vspace*{0.15cm}

As shown in Appendix~\ref{sec:appendix_A}, this result holds as the characteristic function 
\begin{equation}\label{eq:G_Pfi}
    \mathcal{G}(j\eta,t) = \sum_{i,f}P_{i}P_{f|i}(t)e^{-\eta\Delta E_{i,f}(t)}
\end{equation}
is analytic and convex in $\eta$~\cite{Rockafellar70,TouchettePhysRep2009}. In Eq.~(\ref{eq:G_Pfi}), $P_i$ is the probability of measuring $E_i$ at the beginning of the protocol, $P_{f|i}$ is the conditional probability to get $E_f$ after having initially measured $E_i$, and $\Delta E_{i,f}(t) \equiv E_f(t) - E_i(t_0)$. 

We can now establish the following:

\vspace*{0.15cm}
\noindent
{\bf Corollary 1}~[\emph{Application of the Lemma to unital quantum maps}]: If the initial state is thermal at inverse temperature $\beta$ and the open quantum dynamics is such that the asymptotic state is the completely mixed state, then the non-trivial solution of $\mathcal{G}(j\eta^{*},t) = 1$ at $t \rightarrow \infty$ is $\eta^{*} = \beta$.
\vspace*{0.15cm}

The validity of Corollary 1 can be verified by making the substitution $\eta^{*} = \beta$ and $P_{f}(\infty)=1/n$ in $\lim_{t\rightarrow\infty}\mathcal{G}(j\eta^{*},t)$, where $P_{f}(\infty)$ is the probability to measure the $f$-th energy value of the system in the correspondence of the asymptotic state. 

\section{Memory loss assumptions}
\label{sec:memory_loss}

In this section we present a discussions of assumptions under which (\ref{eq:QR_steady_states}) is valid, not only asymptotically ($t \rightarrow \infty$), but for $t \geq t^{*} \geq 0$, still using $\eta^{*}$ as the energy scale factor. The main motivation in understanding whether and how to rescale energy exchange fluctuations over time with $\eta^{*}$, which we recall depends only on the initial and the asymptotic states, lies within the objective to generalize the Jarzynski-W\'ojcik fluctuation theorem to a generic dissipative quantum system. In these terms, $\eta^{*}$ can be considered as a macroscopic quantity that effectively acts as the proxy of an inverse temperature~\cite{schrodinger1989statistical}.

One would like to determine necessary and sufficient condition for having $\mathcal{G}(j\eta^{*},t) = \langle e^{-\eta^{*}\Delta E(t)}\rangle = 1$ for $t \geq t^{*}$ [Eq.~(\ref{eq:QR_steady_states})]. Even finding a necessary condition is a very complicated task in the general case, where no specifications on the (thermo)dynamic process under scrutiny are provided. So, we are going to introduce a sufficient condition that guarantees the validity of Eq.~(\ref{eq:QR_steady_states}) depending on properties (specified below) of the energy exchange statistics. 
In particular, we will focus on dynamics where memory of the initial state is partially lost, which generalizes the case of thermalizing quantum dynamics~\cite{Ramezani18}.
In doing this, we are led by the requirement to work within a dynamical regime that {\it (i)} can be associated to paradigmatic models with a clear physical meaning; and {\it (ii)} can be tested in meaningful experimental scenarios. 

Regarding point {\it i)}, we are going to consider an extended class of Stern-Gerlach protocols where a series of projective measurements is performed sequentially, and among the measurements a dynamical quantum map is included. Moreover, we also give the possibility that each measurement does occur (or fails) with a given probability. Lastly, concerning an experimental scenario where our findings can be tested, we will consider a dissipative quantum Maxwell's demon. 
Thus, our theory will be experimentally tested on a single NV center in diamond at room-temperature that is subjected to a sequence of laser pulses, each of which acts as a combination of a quantum measurement and a dissipative channel. Deviations of the theoretical assumption enabling Eq.~(\ref{eq:QR_steady_states}) from experimental data are analyzed in Section \ref{sec:exp}.

\subsection{Generalized Stern-Gerlach protocols}

\begin{figure*}[t]
\centering
\includegraphics[width=0.9\textwidth]{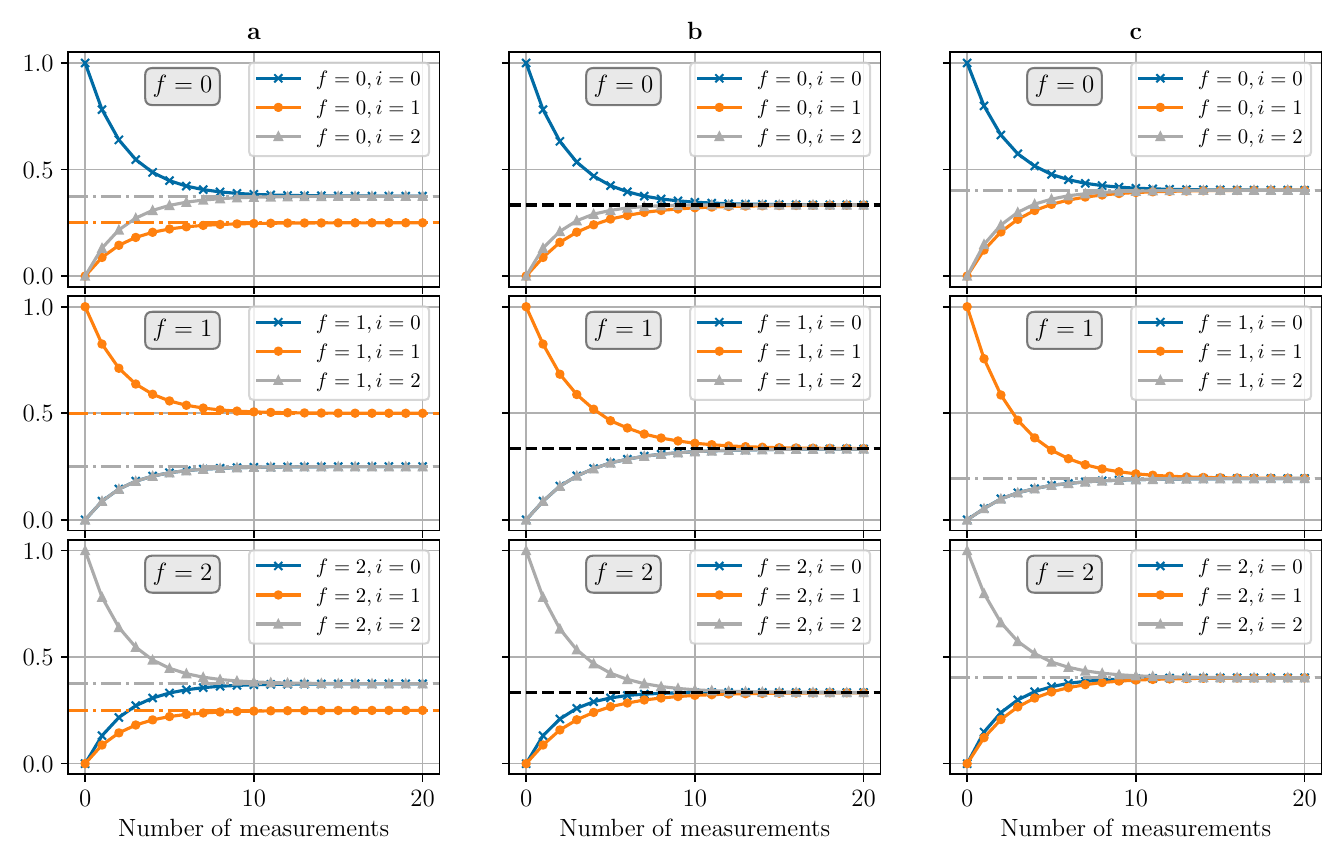}
\caption{
Energy jump conditional probability $P(f|i)$ associated with the Stern-Gerlach-like experiment where the Hamiltonian of the three level system, $\mathcal{H}\propto S_x$, is measured once at the beginning of the protocol and once at the end. In-between these measurements, a set of measurements of the observable $S_z$ occurs with finite probability ($p = 0.35$). The x-axis indicate the number of these intermediate measurements applied to the system. The states $0$ and $2$ correspond to the eigenstates of $S_x$ with positive and negative eigenvalues respectively, while the state $1$ refers to the eigenstate with associated eigenvalue equal to zero. 
{\bf (a)}~No dynamics in-between measurements. The unitary operator describing the evolution of the system is the identity. Here, the state of the system asymptotically approaches the state in the case the probability of absorption is equal to one, which is indicated by the dotted-dashed lines.   
{\bf (b)}~Unitary evolution in-between measurements. The unitary evolution affects the state of the system. The number of different trajectories increases as the power of four with respect to the number of measurements. Eventually it is impossible to have information about the initial state of the system. This asymptotic behavior is that of a complete memory loss, which is equivalent to a thermalization with an infinite temperature bath.
{\bf (c)}~Same as (b), but also with dissipation conditioned on the measurements (see text). In this case, the system eventually looses information of the initial state, but the asymptotic state is not completely mixed. The asymptotic state is indeed an out-of-equilibrium steady-state.}
\label{fig:Stern_Gerlach}
\end{figure*}

In order to find a sufficient condition for (\ref{eq:QR_steady_states}), we start by observing that (\ref{eq:QR_steady_states}) is implied by the loss of memory, starting from a time $t^{*}$, of the chosen initial state. This is what occurs in the Stern-Gerlach experiment and its variations~\cite{SakuraiBook}, when a measurement of a certain observable is followed by other measurements of non-commuting observables, so that the later measurements cancel the information about the initial state.

Motivated by this observation, we start by considering the simplest Stern-Gerlach protocol for our purposes, namely the sequential measurement (via projective measurements) of a two-level quantum system (e.g., a spin-$\frac{1}{2}$) with respect to three observables: the first and the third, the Hamiltonian $\mathcal{H}$ of the system (say, the Pauli matrix $\sigma_x$), and the intermediate observable an operator that is maximally non-commuting with $\mathcal{H}$ (say, $\sigma_z$). 
Only the outcomes of the first and third measurements are collected, giving origin to the joint probability $P_i P_{f|i}(t)$. For this paradigmatic example, where the probability to measure $\sigma_z$ (at time $t^*$) is equal to $1$, the conditional probabilities $P_{f|i}(t)=1/2$ are independent on $i$, for any $f$ and $t \geq t^{*}$. 
In fact, whatever is the initial state, consecutively measuring the Pauli matrices $\sigma_x$ and $\sigma_z$ leads to a {\it complete destruction} of the information contained in the initial state~\cite{SakuraiBook}. 

We would be then tempted to set as a sufficient condition for \eqref{eq:QR_steady_states} the following hypothesis:

\vspace*{0.15cm}
\noindent
{\bf Hypothesis $\mathcal{I}^*$}~[\emph{Complete memory loss of the initial energy statistics from a time $t^*$}]: For any final state $f$, $P_{f|i}(t)$ is independent of all initial states $i$ for $t>t^*$. 
\vspace*{0.15cm}

However, we will argue that the hypothesis $\mathcal{I}^*$ is unnecessarily too restrictive, and possibly unfeasible for quantum systems with dimension larger than 2. Sequential measurements over maximally non-commuting observables lead to complete memory loss, irrespective of the system dimension. In general, determining maximally non-commuting observables that are also experimentally implementable is a non-trivial, sometimes impossible, task. 
Orthogonal observables, such as the angular momentum operators easily obtained in experimental routines, can provide a readily accessible non-commuting set, albeit no  maximal non-commuting. For example, the orthogonal angular momentum operators $S_x$ and $S_z$ are not maximally non-commuting even for a qutrit, a system of relevance to our experiments. Consequently, alternating measurements of $S_x$ and $S_z$ (e.g., measuring $S_x$, $S_z$ and $S_x$) does not completely erase the information about the initial state, and the condition of complete memory loss is no longer valid since $P_{f|i}(t)$ still depends on $i$. The need of finding maximally non-commuting observables can be overcome by introducing a dynamical evolution given by a map $\Phi$ interspersed by consecutive measurements, so as to scramble the state of the system and induce memory loss.

Let us further analyze these cases. Consider a protocol in which the first and last measurements are the same as before---projective measurements of the Hamiltonian (e.g., the angular momentum operator $S_x$)---and we still consider the case where only the outcomes of the first and last measurements are collected. Instead, the second measurement ($S_z$) is replaced by a sequence of projective measurements that occur with a probability $p_m < 1$. These intermediate measurements all refer to the same observable (e.g., $S_z$). Formally, a quantum measurement with an occurrence probability $< 1$ is described by a positive operator valued measure (POVM): with probability $1 - p_m$ the system remains unaltered, and with probability $p_m$ its wave-function collapses due to the effect of a projective measurement (in our example into one of the eigenstates of $S_z$). If no evolution between consecutive measurements is considered, no complete memory loss is attained, as shown in Fig.~\ref{fig:Stern_Gerlach}(a). Complete memory loss is reintroduced if we modify the protocol by including a unitary evolution of the system between the measurements. In fact, if the unitary operator $U$ describing this evolution is equal to the identity, or if it commutes with the intermediate POVMs (in our case when $[U,S_z]=0$), then we go back to the previous case depicted in Fig.~\ref{fig:Stern_Gerlach}(a). In any other case, where $U$ is not the identity and it does not commute with the POVMs, there exists a time $t^{*}$ (or a finite number of intermediate measurements) after which all the conditional probabilities $P_{f|i}(t)$ tend to $1/n$, with $n$ the dimension of the quantum system. This circumstance, shown in Fig.~\ref{fig:Stern_Gerlach}(b), entails the complete destruction of information for $t \geq t^*$, in agreement with previously known results where dynamics leading to infinite-temperature thermalization (i.e., the state of the system converges towards a completely mixed state) can be induced by a sequence of projective measurements under specific conditions~\cite{Yi11,GherardiniPRE2021}.

We are now in the position to analyze the behaviour of the conditional probabilities $P_{f|i}(t)$ in Stern-Gerlach-like experiments that also include {\it dissipation}. We consider a sequence of intermediate projective measurements of $S_z$ interspersed by unitary operators $U$ not commuting with $S_z$. As before, the first and the last energy measurements are performed along the basis of $S_x$, and their outcomes are collected. We still consider that the intermediate measurements occur with a probability $p_m$, but in this case if the measurement occurs then the state of the system collapses and changes due to a dissipative channel. 
In Fig.~\ref{fig:Stern_Gerlach}(c) we show an example of such dynamics that consider a dissipative channel pushing the state of the system into one of the eigenstates of the observable $S_z$. It is worth noting that this case is not purely academic, since it can be used to interpret the effect of \emph{optical pumping} with short excitation pulses in realistic experimental scenarios, where the probability of light absorption is smaller than 1.
An example of this is the implementation of a Maxwell's demon on an NV center via a train of short laser pulses~\cite{HernandezGomezPRXQuantum22}. In these Stern-Gerlach-like experiments with dissipation, the hypothesis of complete destruction of information is not fulfilled, but a less stringent one, which still maintains a symmetry in the indexes of the measurement outcomes, holds. Formally, if the dissipation is present, the following hypothesis, which will be extensively considered in the remaining of the paper, is valid:

\vspace*{0.15cm}
\noindent
{\bf Hypothesis $\mathcal{I}$}~[\emph{Almost complete memory loss of the initial energy statistics}]: For any $f$ and $t \geq t^{*}$, the conditional probabilities $P_{f|i}(t)$ are independent on all the values taken by $i$ except $i=f$, i.e., $i \neq f$.
\vspace*{0.15cm}

Clearly, Hypothesis $\mathcal{I}^*$ implies 
Hypothesis $\mathcal{I}$, but not vice versa.

\section{Fluctuation theorem}
\label{sec:Theorem}

We can now provide the statement of the fluctuation theorem formalizing Eq.~\eqref{eq:QR_steady_states}. For this purpose, we express the conditional probabilities $P_{f|i}$ as a function of the probabilities $P_{f}(\infty)$ in the asymptotic limit (assumed to exist). Specifically, for $i \neq f$, one is always allowed to write the conditional probabilities $P_{f|i}$ as $P_{f|i}(t) = F_{i,f}(t)P_{f}(\infty)$ with $F_{i,f}(t)$ a generic bounded real function depending on the indices $i,f$, such that $F_{i,f}(t_0)=0$ and $\lim_{t \rightarrow \infty}F_{i,f}(t) = 1$ $\forall\,i,f$. Thus,

\vspace*{0.15cm}
\noindent
{\bf Theorem}~[{\it Fluctuation theorem for dissipative quantum dynamics}]: Under the validity of both the hypothesis $\mathcal{I}$ and the detailed balance condition, implying $P_{f|i}(t) = \overline{F}(t)P_{f}(\infty)$ $\forall\, i, f, t \geq t^{*}$ with $f \neq i$ and $\overline{F}(t)$ a time-dependent real function, then
\begin{equation}\label{eq:Theorem}
\mathcal{G}(j\eta^{*})=1 \quad \text{at} \quad t \rightarrow \infty \quad \Longleftrightarrow \quad \mathcal{G}(j\eta^{*})=1 \quad \forall t \geq t^{*} \,.
\end{equation}

\vspace*{0.15cm}
The proof of the Theorem is in Appendix \ref{app:proof_theorem}. It is worth noting that the assumption of the Theorem includes the DBC 
\begin{equation}\label{eq:DBC}
\frac{P_{f|i}(t)}{P_{i|f}(t)} = \frac{P_{f}(\infty)}{P_{i}(\infty)}
\end{equation}
that is obeyed by any reversible Markov process. 
Moreover, since the hypothesis $\mathcal{I}$ is implied by the complete destruction of information, then also assuming the latter leads to the validity of the Theorem's thesis. The Theorem provides a sufficient condition such that $\mathcal{G}(j\eta^{*})=1$ for $t \geq t^{*}$ using the energy scale factor $\eta^{*}$ that only depends on the initial and asymptotic states. The Theorem holds independently of the Hamiltonian, as long as its eigenvalues are time-independent, and of the dimension of the system. Moreover, the Theorem does not require a specific initial state, provided however the TPM scheme is applied. In the limiting case where the initial and asymptotic states are thermal with inverse temperatures $\beta$ and $\beta_\infty$ respectively, the energy scale factor $\eta^{*} = \Delta\beta = \beta - \beta_\infty$. Hence, Eq.~\eqref{eq:QR_steady_states} reduces to the exchange fluctuation relation valid for a quantum system under thermalizing dynamics~\cite{Jarzynski04,Ramezani18}, or under an effective thermalizing quantum map~\cite{HernandezGomez20} as it occurs in two-level quantum systems~\footnote{For two-level quantum systems there always exists an effective temperature such that the elements of any mixed state can be sampled from a Boltzmann distribution. On the contrary, higher-dimensional quantum systems (as in the present paper) provide genuinely quantum states that cannot be mapped into thermal states.}. 

We conclude this section showing that the sign of $\eta^{*}$ determines whether energy is injected or can be extracted from a dissipative system:

\vspace*{0.15cm}
\noindent
{\bf Corollary 2}~[\emph{$\eta^{*}$ as a measure of the energy extraction power}]: At $t$ large ($t \geq t^{*}$), when the quantum system under dissipation has reached the asymptotic state, \emph{necessary and sufficient condition} for \emph{energy extraction} is $\eta^{*}<0$.

\vspace*{0.15cm}
Corollary 2 is obtained by applying the Jensen's inequality on Eq.~\eqref{eq:QR_steady_states} that thus implies $\eta^{*}\langle\Delta E\rangle \geq 0$. As a result, energy is extracted from the system when $\eta^{*}<0$, and is transferred to the system otherwise ($\eta^{*}>0$).

\begin{figure*}[t]
\centering
\includegraphics[width=0.75\linewidth]{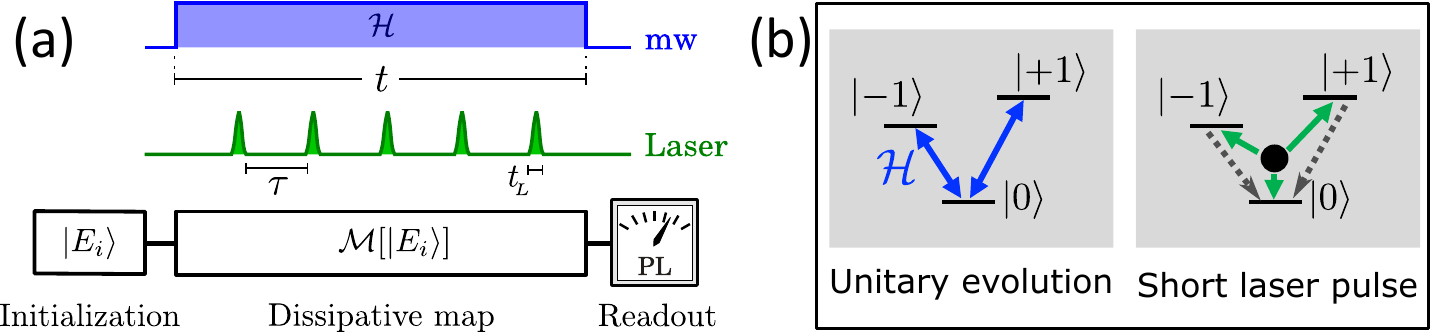}
\caption{
{\bf (a)} Sketch of the experimental protocol: The NV spin qutrit ($S_z = [0, \pm 1]$) is initialized in an eigenstate of the mw Hamiltonian, and evolves under a map $\mathcal{M}$ given by the combination of $\mathcal{H}$ and short laser pulses~\cite{HernandezGomezPRXQuantum22}. After the evolution for a time $t$, the spin energy is measured according to the TPM scheme. {\bf (b)} The map $\mathcal{M}$ operates on the NV through a unitary evolution (generated by $\mathcal{H}$) and a combination of POVMs and dissipation towards the spin state $\ket{0}$ (via laser pulses), sketched in the left and right panels, respectively.}
\label{fig:Fig1}
\end{figure*}

\section{Experimental scenario: Dissipative quantum Maxwell's demons}
\label{sec:exp}

In this section, we introduce an experimental testbed very relevant for our purposes: the dissipative quantum Maxwell's demon, originally introduced in~\cite{HernandezGomezPRXQuantum22}. The quantum maps realized in this physical platform admit a non-equilibrium steady-state that is induced by a sequence of measurements followed by feedback actions. With this setup we test the validity of the Theorem's assumptions, as well as its thesis.

The experimental platform is based on single NV centers, where the NV electronic spin $S = 1$ is optically initialized and read out, and manipulated with a Hamiltonian $\mathcal{H}$ via a microwave (mw) radiation. A detailed discussion of the apparatus, as well as the experimental realization (TPM scheme included) is reported in~\cite{HernandezGomezPRXQuantum22}. 
Here we give a summary of the protocol.  
The NV bare Hamiltonian commutes with the spin operator $S_z$ (with eigenstates $\ket{0}$,$\ket{\pm 1}$). In the presence of a near-resonance mw field, the system Hamiltonian $\mathcal{H}$ can be written in a rotating frame as a time-independent linear combination of the operators $S_x$, $S_y$ and $S_z$. During the unitary evolution determined by this Hamiltonian, the NV spin state, represented by the density operator $\rho$, is subjected to a series of short laser pulses, inducing POVMs and dissipation (see Fig.~\ref{fig:Fig1}). 
In the terminology used in Sec.~\ref{sec:memory_loss}, in the presence of a single laser pulse the NV spin remains unaltered with a probability $1-p_m$, and it is subjected to a projective measurement ($S_z$) and dissipation towards $m_S=0$ with a probability $p_m$. 
A formal model of the NV photodynamics involves radiative and non-radiative transitions, including fast phononic relaxation whose details will not be covered here, since they are thoroughly explained in Ref.~\cite{HernandezGomezPRXQuantum22}. 
The combination of unitary evolution under the Hamiltonian $\mathcal{H}$ and short laser pulses give rise to the CPTP map $\mathcal{M}$. Notably, the quantum dynamical map $\mathcal{M}$, applied to the $\mathcal{H}$ eigenstates, leads the system to a non-equilibrium steady-state originated by a non-trivial interplay between quantum measurements and dissipation. The described experiment embodies the right tool to demonstrate the robustness of the theorem presented in this work. In fact, the possibility to design different Hamiltonian operators for the unitary evolution of the spin-qutrit allows us to implement various protocols where the DBC may be fulfilled or not.

We would stress that this case-study is appropriate for the test of the fluctuation theorem introduced in Sec.~\ref{sec:Theorem}, since the Hamiltonian of this quantum system can be controlled such that the hypothesis $\mathcal{I}$ of the Theorem as well as the DBC, leading to the assumption $P_{f|i}(t) = \overline{F}(t)P_{f}(\infty)$ for $t \geq t^{*}$, are valid. Specifically, when $[\mathcal{H}, S_z]=0$ the Theorem is fulfilled. In contrast, when $\mathcal{H}\propto S_x$, the hypothesis $\mathcal{I}$ is only approximately valid, but still $\mathcal{G}(j\eta^{*}) \simeq 1$ \emph{within experimental precision} for $t \geq t^{*}$. Even in this case where the Theorem's assumption $P_{f|i}(t) = \overline{F}(t)P_{f}(\infty)$ holds only approximately or is even not valid, we can experimentally explore the deviations from the predictions of the Theorem. 

\subsection{Robustness analysis}

\begin{figure}[t]
\centering
\includegraphics[width=0.975\columnwidth]{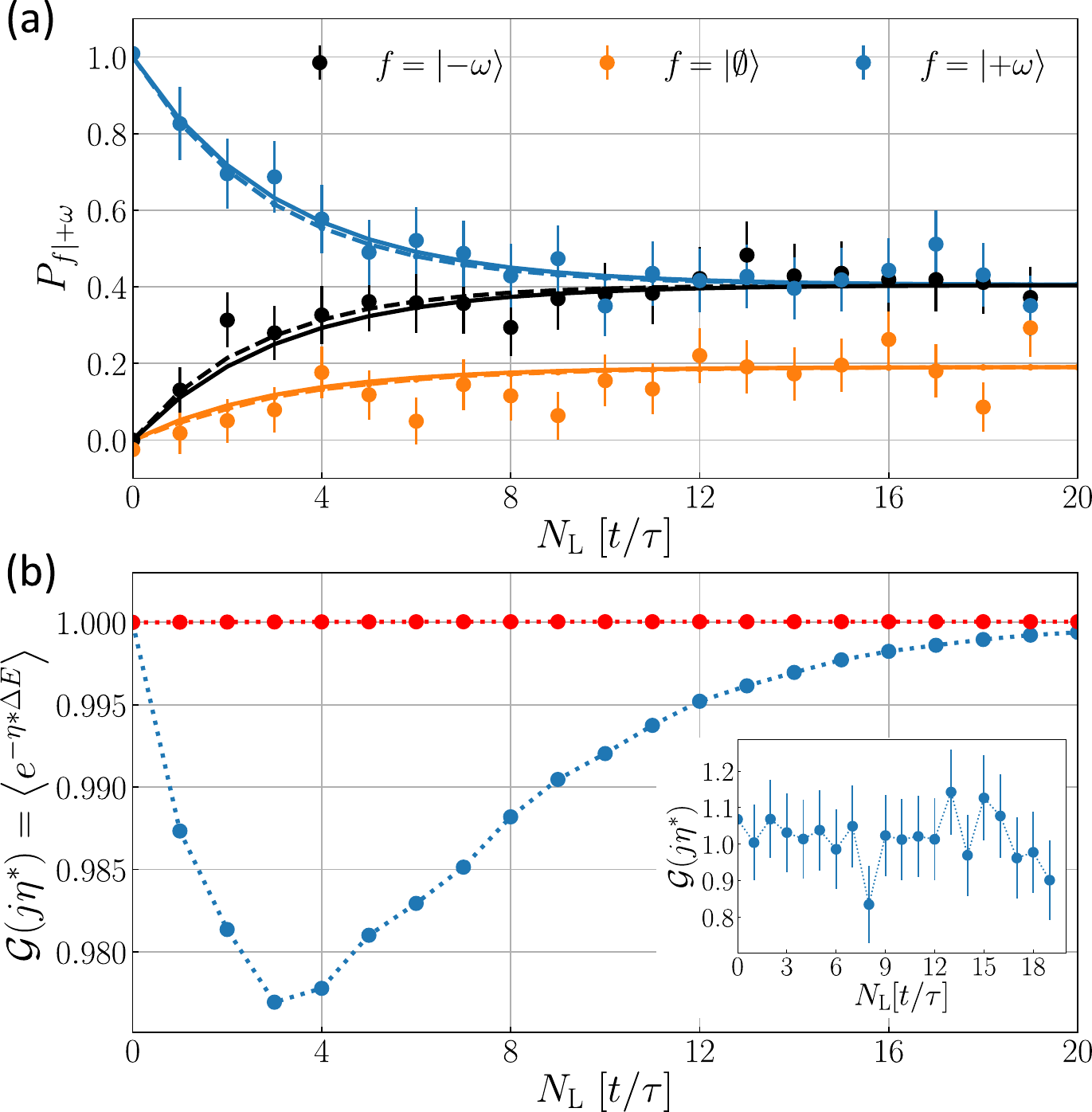}
\caption{
{\bf (a)}: Conditional probabilities of measuring $\ket{E}_f \in \{ \ket{-\omega}, \ket{\emptyset}, \ket{+\omega} \}$ while initializing the qutrit in $\ket{+\omega}$, as a function of the number of laser pulses $N_\mathrm{L} = t/\tau$, with $t$ equal to $\mathcal{M}$ duration and $\tau$ the time interval between pulses. Colored scatters and dashed lines correspond to experimental data and numerical simulations, respectively, of the NV spin evolution under the map $\mathcal{M}$, with $\mathcal{H} = \omega S_x$. Solid lines represent the approximated model  ensuring the DBC, obtained by fitting the numerical results using the function $P_{f|+\omega} = P_f(\infty) (1-e^{-t/\tau_{D}}) + \delta_{+\omega,f} \, e^{-t/\tau_{D}}$, with $\delta_{+\omega,f}$ Kronecker delta and $\tau_{D}=3.11 \tau$. 
{\bf (b)}: Characteristic function $\mathcal{G}(j\eta^{*})$ as a function of the number of laser pulses $N_\mathrm{L} = t/\tau$. $\mathcal{G}(j\eta^{*})$ is evaluated from theoretical simulations (main plot) and experimental data (inset) by considering all the possible combinations $P_{f|i}$, with $i,f \in \{-\omega, 0, +\omega\}$ (see~\cite{HernandezGomezPRXQuantum22}). The comparison between the approximated model with the DBC (in red) and the original one (blue) highlights a discrepancy, as expected from the Theorem.}
\label{fig:Fig2}
\end{figure}

This section is devoted of investigating how much $\mathcal{G}(j\eta^{*})$ deviates from $1$ in case the assumptions of the Theorem are not fulfilled. A possible cause for such a condition is that the DBC (\ref{eq:DBC}) does not hold, which in our case-study occurs by setting $\mathcal{H} = \omega S_x$. Here, $S_x$ is the spin operator orthogonal to the natural NV quantization axis $z$, and $\omega$ is the driving Rabi frequency. The energy eigenstates are $\ket{E}_f \in \{ \ket{-\omega}, \ket{\emptyset}, \ket{+\omega} \}$, with $|\pm \omega\rangle \equiv (\ket{-1} \pm \sqrt{2} \ket{0} + \ket{1})/2$ and $\ket{\emptyset} \equiv (\ket{-1} - \ket{1})/\sqrt{2}$. In Fig.~\ref{fig:Fig2}(a) we report the conditional probabilities $P_{f|+\omega}$ of measuring $\ket{E}_f$ as final states of the TPM scheme (at time $t$, after $N_\mathrm{L}$ laser pulses) once the NV spin is initialized in $\ket{+\omega}$. First, we find a good agreement between the experimental data (colored dots) with theoretical simulations of the evolution under the quantum dissipative model (dashed lines). More interestingly, by imposing $P_{f|i} = P_f(\infty) (1-e^{-t/\tau_{D}}) + \delta_{i,f} \, e^{-t/\tau_{D}}$ with $\delta_{i,f}$ denoting the Kronecker delta between $i$ and $f$, $\tau_{D}$ a decay time and $i = +\omega$ in Fig.~\ref{fig:Fig2}, it is possible to approximate the time evolution of the system to a case where the assumption $P_{f|i}(t) = \overline{F}(t)P_{f}(\infty)$ holds. Despite the validity of this assumption has to be ascribed to a quite different physical scenario, even in this case the agreement between simulations (solid lines in Fig.~\ref{fig:Fig2}(a)) and experiments is still in place. Then, by comparing the characteristic functions of the two models for the corresponding $\eta^{*}$ value (Fig.~\ref{fig:Fig2}(b)), we observe that $\mathcal{G}(j\eta^{*})=1$ only for the detailed balanced case, as expected from the Theorem. On the other hand, their difference is much smaller than our experimental uncertainty (see inset of Fig.~\ref{fig:Fig2}(b)), proving the robustness of the fluctuation relation (\ref{eq:QR_steady_states}) against the possible experimental invalidation of the assumptions.

\subsection{Dependence of $\eta^{*}$ on the values of the Hamiltonian parameters}

\begin{figure}[t]
\centering
\includegraphics[width=0.95\columnwidth]{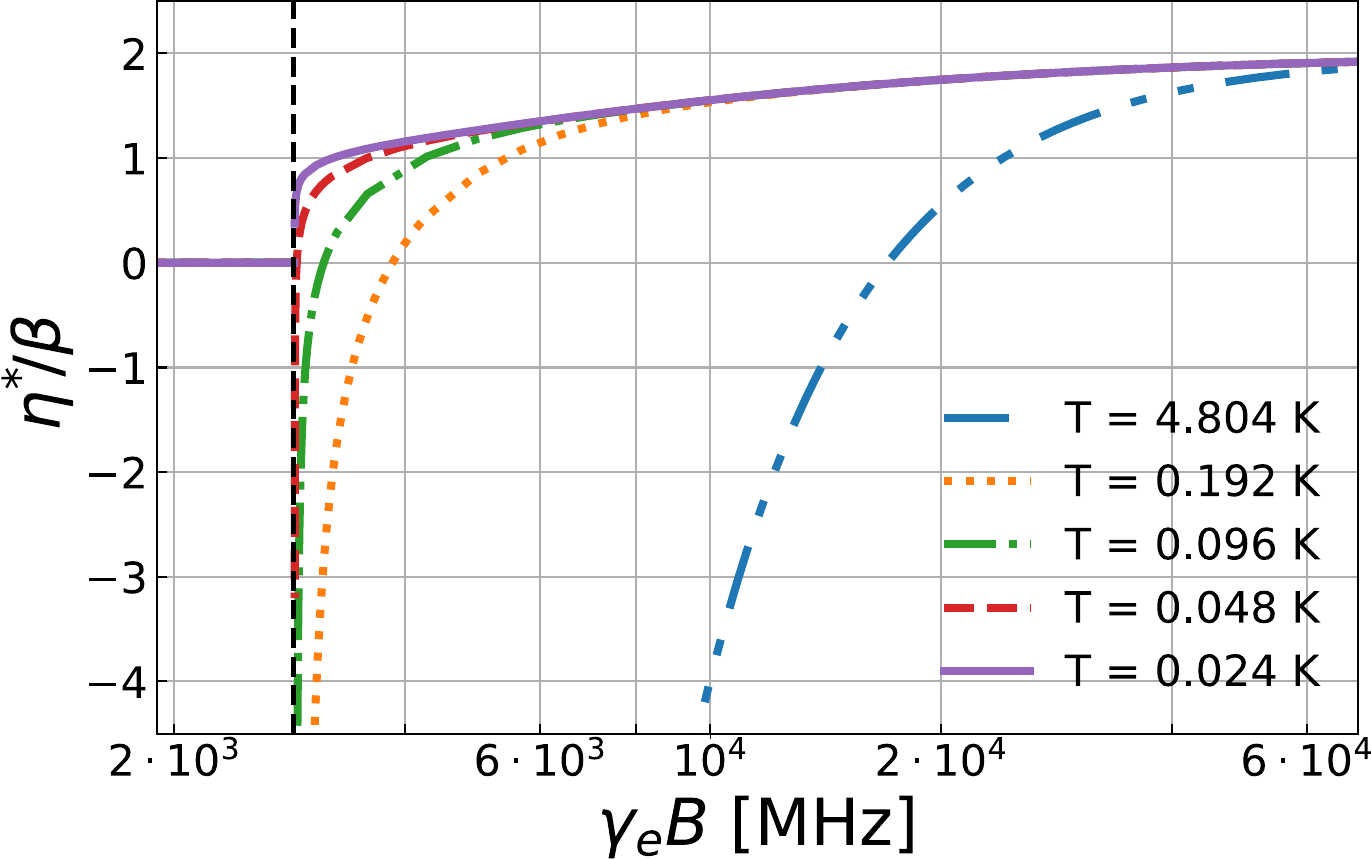}
\caption{
Ratio between the energy scale factor $\eta^{*}$ and the initial temperature $T$ as a function of $\gamma_e B$. Lines are obtained by directly solving the characteristic function $\mathcal{G}$ (as discussed in detail in Appendix~\ref{sec:appendix_C}) through a standard optimization routine. At the $\ket{-1} \leftrightarrow \ket{0}$ energy crossing ($\gamma_e B = \Delta$) we find a discontinuity in $\eta^{*}$.
}
\label{fig:Fig3}
\end{figure} 

We consider now an experimental case where the assumptions used to obtain the Theorem holds. This is obtained by turning off the mw radiation and using the natural NV spin Hamiltonian $\mathcal{H} = \Delta S_z^2 + \gamma_e B S_z$, with $\Delta$ the zero-field splitting, $\gamma_e$ the electronic gyromagnetic ratio, and $B$ a bias external magnetic field aligned with the NV quantization axis. Here, the effect of the laser pulses train, always pumping the NV spin in $\ket{0}$, is to make the qutrit collapse in one of the energy eigenstates. This allows us to find a simplified analytical solution to the characteristic function $\mathcal{G}$ with respect to the values of the Hamiltonian parameters (see Appendix~\ref{sec:appendix_C}).

We can now distinguish two different regimes: $\gamma_e B < \Delta$, and $\gamma_e B > \Delta$, shown in Fig.~\ref{fig:Fig3}. Here, for the sake of simplicity, we considered initial thermal states, with $P_i = e^{-\beta E_i}/\sum_k e^{-\beta E_k}$, $k = {1,\ldots,n}$. In the first regime, the energy dissipation leads the system in the state corresponding to the energy minimum $E_{\ket{0}} = 0$. This means that the asymptotic energy level is related to a thermal state at zero temperature, $\beta_{\infty} = \infty$. As a consequence, there is only a trivial solution solving the fluctuation theorem, that is, $\eta^{*} = 0$.

In the second case ($\gamma_e B > \Delta$), the eigenstate associated to the minimum energy of the system is $\ket{-1}$, and we observe a \emph{non-linear} relation of $\eta^{*}$ as a function of $\beta$, meaning that the quantity $\eta^{*} - \beta$ in general cannot be associated with an inverse temperature. Notably, instead, for a magnetic field $B \rightarrow \infty$ the value of $\eta^{*}$ becomes constant and is equal to $2\beta$. In this way, by knowing the temperature of the initial thermal state, i.e., the partition function at time $t = 0$, we can evaluate $\eta^{*}$ and reconstruct the whole dynamics of the system. This is equivalent to the case with symmetric energy levels, whose analytic solution is studied in Appendix~\ref{sec:appendix_C}. Finally, the special condition $\gamma_e B \rightarrow \Delta^{+}$ leads to a degeneracy of $\ket{0}$ and $\ket{-1}$ eigenstates. In this case, $\eta^{*}$ shows a discontinuity at $-\infty$ that occurs in correspondence of the energy crossing between $E_{\ket{-1}}$ and $E_{\ket{0}}$ (see Appendix~\ref{sec:appendix_C} for more details). 

\section{Conclusions}\label{sec:conclusions}

In this paper, we discussed how to characterize energy exchange statistics in open quantum dynamics (both unital and non-unital) exhibiting dissipation that brings the system towards an asymptotic state not necessarily thermal. In doing this, we derive a quantum fluctuation theorem for a family of parameterized dynamics that go beyond thermalizing dynamics, hence extending the results by Jarzynski-W\'ojcik~\cite{Jarzynski04}. First of all, we show that in the asymptotic limit ($t\rightarrow\infty$) the characteristic function $\mathcal{G}(j\eta)$ of the exchanged energy distribution is independent on the process details and is equal to $1$ for a unique, constant (time-independent), non-trivial energy scale factor $\eta=\eta^{*}$. In addition, we show a sufficient condition under which $\mathcal{G}(j\eta^{*})=1$ is valid for any time $t \geq t^{*}$, thus providing us a truly fluctuation theorem even for the quantum dissipative case.

The condition that allows for $\mathcal{G}(j\eta^{*})=1$ for $t \geq t^{*}$ is $P_{f|i}(t) = \overline{F}(t)P_{f}(\infty)$, which is implied by the validity of the hypothesis $\mathcal{I}$ and the detailed balance condition (DBC). As shown in Sec.~\ref{sec:memory_loss}, the hypothesis $\mathcal{I}$ is related to the almost complete memory loss of the initial energy statistics, and can be obtained (and thus tested) in dissipative instances of Stern-Gerlach protocols. According to the hypothesis $\mathcal{I}$, the conditional probabilities of the exchanged energy distribution have to be uniquely determined by the asymptotic probabilities $P_{f}(\infty)$ that enter the non-equilibrium state $\lim_{t\to\infty}\Phi_t[\rho_0]$ reached asymptotically by a quantum system under dissipation. Also the DBC is a necessary requirement for $P_{f|i}(t) = \overline{F}(t)P_{f}(\infty)$. In fact, as shown in Appendix~\ref{app:proof_theorem}, given $P_{f|i}(t) = \widehat{F}_{f}(t)P_{f}(\infty)$ (implied by the hypothesis $\mathcal{I}$), then the DBC entails $\widehat{F}_{f}(t)=\overline{F}(t)$.

Experimentally, we took advantage of a platform based on single NV centers to test the robustness of the Theorem against the fulfillment of its assumptions. While numerical simulations show a discrepancy between the cases with or without the DBC, the experimental outcomes cannot distinguish them within our statistical uncertainty. This outlines that the fluctuation theorem is able to describe systems even beyond its hypotheses. Moreover, by exploring the behaviour of $\eta^{*}$ depending on the value of $\gamma_e B$, we observe a non-linear dependence of $\eta^{*}$ from the initial state, and the presence of a discontinuity in correspondence of an energy level crossing.

As outlook, studies on the energy statistics and tests of the fluctuation theorem at the NV ground state level anti-crossing ($\gamma_e B = \Delta$) would highlight the relationship between the thermodynamic properties of the system and its physical counterparts, as the NV-photoluminescence~\cite{Ivady21}. It is also worth investigating a tighter condition, still sufficient in case, under which the fluctuation theorem (\ref{eq:QR_steady_states}) is valid. Moreover, this work opens the possibility to evaluate the energetics of stabilizing a qubit by means of a dissipative protocol~\cite{Lu17} even with other experimental platforms, or to associate a unique energy scale factor to a quantum many-body phases embodied in the local energy statistics of an array of qubits~\cite{Ma19,Dutta21}. An important application of that would be the dissipative charging of a (many-body) quantum battery~\cite{BarraPRL2019,ReviewQBattery}. Finally, in the case the initial state and the Hamiltonian are not commuting, other protocols than the TPM scheme can be used to characterize energy fluctuations in the quantum regime~\cite{AllahverdyanPRE2014,DeffnerPRE2016,Micadei20,Gherardini20x,Levy20,LostaglioKirkwood2022,FrancicaPRE2023,SantiniPRB2023,Gherardini2024Tutorial}. By applying these methodologies, our results might be no longer valid and some `quantum corrections' may be needed as in~\cite{BelliniPRL2022}. One would thus extend some results in~\cite{Scarani02} that concern the interplay of quantum and classical information processes in dissipative dynamics.

\section*{Acknowledgements}
The authors warmly thank Michele Campisi for collaboration in the initial stage of the work, for a critical reading of the draft and for providing us useful suggestions. 
A.T.~and S.G.~acknowledge the kind hospitality at the Massachusetts Institute of Technology (MIT) in Boston, where main discussions were carried out.
The authors acknowledge financial support from the MISTI Global Seed Funds MIT-FVG Collaboration Grants ``NV-centers for the Quantum Jarzynski Equality (NVQJE)'' and ``Non-Equilibrium Thermodynamics of Dissipative Quantum Systems (NETDQS)'', the European Commission-EU under GA n.~101070546--MUQUABIS, and the European Union--Next Generation EU within the PNRR MUR project PE0000023-NQSTI.

\onecolumngrid
\appendix

\section{Proof of the Lemma}
\label{sec:appendix_A}

The characteristic function $\mathcal{G}(j\eta)$ of the distribution for the internal energy variation is convex with respect to the scaling factor $\eta$~\cite{Rockafellar70,TouchettePhysRep2009}. Moreover, being given by the finite sum of analytical functions (exponentials), for finite-dimensional systems $\mathcal{G}(j\eta)$ is also a twice differentiable real-valued function with positive concavity. As first step of the proof, our aim is to demonstrate that, at the asymptotic limit $t\rightarrow\infty$, the equality $\mathcal{G}(j\eta)=1$ holds only for $\eta_0 = 0$ [trivial (constant) solution] and for another energy scale factor $\eta^{*}$ that, in the general case, is a real number $\neq 0$, except a limit case whereby $\eta^{*}$ does not exist.\\
To prove this, one can verify the validity of the following two conditions: (i) $\eta_0 = 0$ is a zero of the function $g(\eta) \equiv \mathcal{G}(j\eta) - 1$ s.t.~$g(\eta_0)=0$; (ii) the unique minimum of $g(\eta)$ (in fact, even $g(\eta)$ is a convex function) is equal or smaller than zero. In this way, if both these conditions hold, then $g(\eta)$ can have only another zero $\eta^{*} \neq 0$ at the generic time $t$, since the functions $\mathcal{G}(j\eta)$ and $g(\eta)$ have positive concavity. In the limit cases (analyzed at the end of the proof) $g(\eta)$ exhibits only the single (trivial) zero $\eta_0 = 0$. \\
Condition (i) can be easily verified, as $\eta_0 = 0$ is the trivial solution of the fluctuation relation $\mathcal{G}(j\eta)=1$. Instead, to verify the validity of condition (ii), one has to check if the derivative $\partial g(\eta)/\partial\eta$ evaluated in $\eta=0$ (i.e., the slope of $g(\eta)$ in $\eta=0$) is different from $0$ for any value of the system parameters, except for the limit case of $\eta_0$ unique zero of $g(\eta)$:
\begin{equation}\label{eq:first_derivative_g}
\left.\frac{\partial g(\eta)}{\partial\eta}\right|_{\eta=0} = -\langle E_{\infty}\rangle + \langle E_{\rm in}\rangle \,,
\end{equation}
where $\langle E_{\rm in}\rangle \equiv \sum_{i}P_{i}E_{i}$ and $\langle E_{\infty}\rangle = \lim_{t \rightarrow \infty} \sum_{f}P_{f}(t)E_{f} = \sum_{f}P_{f}(\infty)E_{f}$ with $P_{f}(t)=\sum_{i}P_{f|i}(t)P_i$. Hence, in the asymptotic limit $t\rightarrow\infty$, the right-hand-side of Eq.~(\ref{eq:first_derivative_g}) is $0$ if and only if $\langle E_{\infty}\rangle = \langle E_{\rm in}\rangle$, namely if the initial and asymptotic energy values are the same. In all the other cases, the right-hand-side of Eq.~(\ref{eq:first_derivative_g}) is different from $0$. This can happen e.g.~when the asymptotic state reached by the system is stationary thermal with the same temperature $\beta_{\infty}$ than the initial thermal state, i.e., $\beta_{\infty}=\beta$. Till now, the derivation allows us to state that the minimum value of $\mathcal{G}(j\eta)$ is necessarily equal or smaller than $1$. This implies the existence of a unique value of $\eta$ different from zero obeying the relation $\mathcal{G}(j\eta)=1$ in the asymptotic limit $t\to\infty$. Before proceeding, it is worth observing that if one would define $h(\eta,\alpha) \equiv \mathcal{G}(j\eta)-\alpha$, with $\alpha$ being a real and positive number not necessarily equal to $1$, the value $\eta=0$ is not longer solution of $h(\eta,\alpha)=0$. As $h(\eta,\alpha)=0$ has only two solutions in $\eta$, just $\alpha=1$ guarantees the existence of a unique non-trivial solution $\eta^{*}$. Therefore, $\alpha = 1$ is the appropriate physical choice for exchange fluctuation relations in open quantum systems that however does not entail any free-energy variation.

Let us now focus on the second step of the proof. We show that if the probabilities to measure the energies (eigenvalues of the Hamiltonian) of the system do not depend on time, nor on the initial state $\rho_0$, then the unique finite zero of $\mathcal{G}(j\eta)$, as well as $g(\eta)$, only depends on the initial and asymptotic states. Such assumption is attained when the state of the quantum system, solution of the dynamical equation of motion, has constant energy and does not depend on $\rho_0$. This dynamical regime can be easily translated in a property of the characteristic function $\mathcal{G}(j\eta)$. For this purpose, note that, in correspondence of the asymptotic state, the conditional probabilities $P_{f|i}$ are invariant with respect to the initially measured energy value, with the result that
\begin{equation}\label{eq_stationary}
P_{f|1} = P_{f|2} = \ldots = P_{f|n} \equiv P_{f}(\infty)
\end{equation}
for any $f$, with $n$ denoting the dimension of the system. Thus, by assuming the validity of Eq.~(\ref{eq_stationary}), $\mathcal{G}(j\eta)$ admits the decomposition
\begin{equation}\label{eq:decomposition_RHS_app}
\mathcal{G}(j\eta)
= \sum_{i}P_{i}e^{\eta E_i}\sum_{f}P_{f}(\infty)e^{-\eta E_f}
= {\rm Tr}\left[e^{\eta \mathcal{H}}\rho_{0}\right]{\rm Tr}\left[e^{-\eta \mathcal{H}}\rho_{\infty}\right],
\end{equation}
where $\rho_{\infty}$ is the asymptotic state of the quantum system. The right-hand-side of Eq.~(\ref{eq:decomposition_RHS_app}) is still a convex function with respect to $\eta$. Hence, there exists a unique value of $\eta$ different from zero, namely $\eta^{*}$, that verifies the equality $\mathcal{G}(j\eta^{*})=1$. Moreover, from the decomposition of Eq.~(\ref{eq:decomposition_RHS_app}), it can be observed that $\eta^{*}$ depends only on the set of probabilities $\{P_{i}\}$ and $\{P_{f}(\infty)\}$, corresponding respectively to the initial and asymptotic states, and on the system energy values.

To conclude the proof, let us consider the limit case whereby $g(\eta)$ (that is always a convex function) is provided by an exponential function with a single zero in $\eta_0=0$. Hence, $\eta^{*}$ does not exist. This circumstance, however, is not in conflict with the two steps of the proof: The only requirements for the latter are {\it i)} having $g(\eta) = 0$ for $\eta=0$ and {\it ii)} the (unique) minimum of $g(\eta)$ is $\leq 0$. Both these requirements are fulfilled by the limit case (in fact, in such a case, the minimum of $g(\eta)$ tends to $-\infty$). We stress that in continuous quantum systems, the limit case is unlikely to occur, as opposed to finite-dimensional systems. As example, we consider the dissipative quantum Maxwell's demon addressed in the main text. For the case where the Theorem holds ($\mathcal{H} \propto S_z$), the characteristic function $\mathcal{G}$ reduces to Eq.~\eqref{eq:simplified_G}. Thus, by initializing the quantum system in a state where either $P_{\ket{+1}}$ or $P_{\ket{-1}}$ are equal to zero, the left-hand-side of Eq.~\eqref{eq:simplified_G} is a simple exponential function. As a consequence, $\eta_0=0$ turns out to be the only solution for $\mathcal{G}=1$.

\section{Proof of the Theorem}\label{app:proof_theorem}

Let us set the context for the proof. We consider a dissipative quantum system, whose time-evolution is governed by a generic open quantum map $\mathcal{E}_{t}$ that admits at least one (non-thermal) map fixed-point. Then, the Hamiltonian $\mathcal{H}=\sum_{k}E_{k}\Pi_{k}$ of the system is such that the eigenvalues $E_{k}$ are time-independent. Thus, for $t\to\infty$ the probabilities to measure the energies (eigenvalues of $\mathcal{H}$) of the system does not depend on time (they are indeed constant values) nor on the initial state $\rho_0$. We denote such probabilities as $P_f(\infty)$. Moreover, the fluctuations of energy variations are evaluated by means of the two-point measurement (TPM) scheme. Finally, for our convenience we express the conditional probabilities of the TPM distribution for the internal energy variation as a function of the asymptotic probabilities $P_f(\infty)$. Specifically, for $i \neq f$ [with $i,f$ the indices over the initial and final (at time $t$ where the TPM scheme is stopped) energies respectively], the conditional probabilities $P_{f|i}(t)$ can be written as 
\begin{equation}\label{eq:assumption_1}
    P_{f|i}(t)= F_{i,f}(t)P_{f}(\infty)
\end{equation}
with $F_{i,f}(t)$ generic bounded real function depending on the indices $i,f$ such that $F_{i,f}(t=0)=0$ and $F_{i,f}(t\rightarrow\infty)=1$ $\forall\,i,f$. We note that, since $P_{i|i}(t) = 1-\sum_{f\neq i} P_{f|i}(t) = 1-\sum_{f\neq i}F_{i,f}(t)P_f(\infty)$ (valid for any $i$), one can generally write
\begin{equation}\label{eq:pf_cond_i_if_pair}
    P_{f|i}(t)=F_{i,f}(t)P_f(\infty)(1-\delta_{i,f})+\left(1-\sum_{f\neq i}F_{i,f}(t)P_f(\infty)\right)\delta_{i,f},
\end{equation}
where $\delta_{i,f}$ denotes the Kronecker delta. Eq.~(\ref{eq:pf_cond_i_if_pair}) holds for any pair $(i,f)$ of indices and time $t$.

We are now in the position to start the proof of the Theorem. For this purpose, we take the characteristic function of the statistics for the energy variations (obtained from applying the TPM scheme):
\begin{equation}
    \mathcal{G}(j\eta) = \sum_{i,f}e^{-\eta(E_f - E_i)}P_{i}\,P_{f|i}(t)
\end{equation}
where $P_i \equiv {\rm Tr}[\rho_{0}\Pi_i]$, $P_{f|i}(t) \equiv {\rm Tr}[\Pi_{f}\mathcal{E}_t(\Pi_i)]$. By using the identity $\sum_{f}P_{f|i}(t)=1$ for any $t$ such that $P_{i|i}(t)=1-\sum_{f \neq i}P_{f|i}(t)$, one obtains
\begin{eqnarray}\label{eq:G_eta}
\mathcal{G}(j\eta) &=& \sum_{i} \left[ P_{i}\left(1-\sum_{f \neq i}P_{f|i}(t)\right) + \sum_{f\neq i} P_{i}\,P_{f|i}(t)\, e^{-\eta(E_f - E_i)} \right] \nonumber \\ 
&=& 1 + \sum_{i}\sum_{f \neq i} P_{i}\,P_{f|i}(t)\left( e^{-\eta(E_f - E_i)} - 1 \right).
\end{eqnarray}
Hence, by substituting (\ref{eq:assumption_1}) in (\ref{eq:G_eta}), we find that
\begin{equation}
    \mathcal{G}(j\eta) = 1 + \sum_{i}\sum_{f \neq i} F_{i,f}(t)\,P_{i}\,P_{f}(\infty)\left( e^{-\eta(E_f - E_i)} - 1\right).
\end{equation}
Let us then apply the assumptions of the Theorem, namely the validity of both the hypothesis $\mathcal{I}$ and the \emph{detailed balance equation}. The first hypothesis means that the functions $F_{i,f}(t)$ do not depend on the index $i$ for $f \neq i$ and $t \geq t^{*}$, i.e.,
\begin{equation}\label{eq_app:hypothesis_I}
    P_{f|i}(t) = \widehat{F}_{f}(t)P_{f}(\infty)\,,
\end{equation}
which, together with the detailed balance condition 
\begin{equation}\label{eq:DBE}
    \frac{P_{f|i}(t)}{P_{i|f}(t)} = \frac{P_{f}(\infty)}{P_{i}(\infty)}\,,
\end{equation}
implies in turn that
\begin{equation}\label{eq:theo_assumption}
P_{f|i}(t) = \overline{F}(t)P_{f}(\infty)
\end{equation}
$\forall\,i, f$ with $f \neq i$ and $t \geq t^{*}$, where $\overline{F}(t)$ is a time-dependent real function. As a result,
\begin{equation}\label{eq:after_theorem_assumption}
    \mathcal{G}(j\eta) = 1 + \overline{F}(t)\sum_{i}\sum_{f \neq i} P_{i}\,P_{f}(\infty)\left( e^{-\eta(E_f - E_i)} - 1\right).
\end{equation}

Now some remarks are in order about the interpretation of the Theorem's assumption (\ref{eq:theo_assumption}). The detailed balance equation \eqref{eq:DBE} is obeyed by any reversible Markov process, where reversibility here has to be meant with respect to the space of events that define the Markov process of energy outcomes. Instead, the hypothesis $\mathcal{I}$ (i.e., assumption \eqref{eq_app:hypothesis_I}) stems from the almost complete loss of memory of the initial state, as discussed in the main text, Sec.~\ref{sec:memory_loss}.

The last step of the proof is to recall the definition of $\eta^{*}$ from the Lemma's thesis. From the Lemma, indeed, we already know that at $t \rightarrow \infty$ $\eta^{*}$ is defined as the energy scale factor such that
\begin{equation}\label{eq:def_energy_scaling_factor}
\lim_{t \rightarrow \infty}\mathcal{G}(j\eta^{*})=1 \quad \Longleftrightarrow \quad \sum_{i}\sum_{f \neq i} P_{i}\,P_{f}(\infty)\left( e^{-\eta^{*}(E_f - E_i)} - 1 \right)=0 \,.
\end{equation}
As a direct consequence, by substituting (\ref{eq:def_energy_scaling_factor}) in (\ref{eq:after_theorem_assumption}), we determine $\mathcal{G}(j\eta^{*})=1$ for $t \geq t^{*}$. This concludes the proof of the thesis of the Theorem in the main text.

\section{Dissipative Maxwell's demon}
\label{sec:appendix_C}

Let us consider a generic finite dimensional quantum system. By diagonalizing the system Hamiltonian at time $t=0$ as $\mathcal{H}
=\sum_{k}E_{k}|E_{k}\rangle\!\langle E_{k}|$, the initial state can be written as $\rho_0 = \sum_{k,\ell}\rho_{k\ell}|E_{k}\rangle\!\langle E_{\ell}|$ where $\rho_{k\ell} \equiv \langle E_{k}|\rho_0|E_{\ell}\rangle$.

Then, we take again the characteristic function of the TPM probability distribution ${\rm P}(\Delta E(t))$, i.e., $\mathcal{G}(j\eta) = \langle \exp(-\eta\Delta E)\rangle = \sum_{i,f}P_{i}P_{f|i}(t)e^{-\eta(E_f - E_i)}$, which depends on the energy scale factor $\eta$.
In the following, we provide the theoretical analysis carried out to test the validity of the Theorem in some meaningful cases of the dissipative quantum Maxwell's demon shown in the main text. First, we consider a case-study with symmetric energy levels of a qutrit, which is applicable to the NV center case for $\mathcal{H}\propto S_z$ and a value of the magnetic field inducing a Zeeman shift $\gamma_e B$ much larger than the so-called ground state level anti-crossing (GSLAC). Notably, this can describe also other exemplary cases, e.g.~the ground state of $^{87}Rb$ atom with $F=1$. Secondly, we explicitly study the case of a NV center relaxing the hypothesis of $\gamma_e B \ll \Delta$, while maintaining $\mathcal{H}\propto S_z$.

We consider that the qutrit has reached the asymptotic regime such that $P_{f|i}(t)=P_{f}(\infty)$. Hence, the expression of $\mathcal{G}(j\eta^{*})=1$ for a generic qutrit is
\begin{eqnarray}\label{eq:app_G_steady_state}
& P_1 P_3(\infty) e^{-\eta^{*}(E_3-E_1)} + P_1 P_2(\infty) e^{-\eta^{*}(E_2-E_1)} + P_1 P_1(\infty) + P_2 P_3(\infty) e^{-\eta^{*} (E_3-E_2)} &\nonumber \\ 
& + P_2 P_2(\infty) + P_2 P_1(\infty) e^{-\eta^{*}(E_1-E_2)} + P_3 P_3(\infty) + P_3 P_2(\infty) e^{-\eta^{*}(E_2-E_3)} + P_3 P_1(\infty) e^{-\eta^{*}(E_1-E_3)} = 1 &
\end{eqnarray}
where $P_i$ and $P_f(\infty)$ are the initial and final asymptotic probabilities respectively.
Thus, if we assume that $E_1=0$, then
\begin{eqnarray}
&P_1 P_3(\infty) e^{-\eta^{*} E_3} + P_1 P_2(\infty) e^{-\eta^{*}E_2} + P_1 P_1(\infty) + P_2 P_3(\infty) e^{-\eta^{*}(E_3-E_2)}&\nonumber \\ 
&+ P_2 P_2(\infty) + P_2 P_1(\infty) e^{\eta^{*} E_2} + P_3 P_3(\infty) + P_3 P_2(\infty) e^{-\eta^{*} (E_2-E_3)} + P_3 P_1(\infty) e^{\eta^{*} E_3} = 1 \,.&
\end{eqnarray}

\subsection{Qutrit with symmetric energies}

Assuming that the energy values of the qutrit are symmetric around zero, we define $E_1=0$, $E_2=-\overline{E}$ and $E_3 = \overline{E}$ with $\overline{E}=\hbar\omega/2$. In this way, by means of the substitution
\begin{equation}
x \equiv e^{-\eta^* \overline{E}} \Longleftrightarrow \eta^* = -\frac{1}{\overline{E}}\ln x \,,
\end{equation}
the equation $\sum_{i,f}P_{i}P_{f}(\infty)e^{-\eta^*(E_f - E_i)} = 1$ can be rewritten as a polynomial equation in $x$. For a qutrit (see Eq.~(\ref{eq:app_G_steady_state})) the polynomial equation is:
\begin{eqnarray*}\label{eq:receipe}
&(x-1)\left[ P_{2}P_{3}(\infty)\,x^{3} + ( P_{1}P_{3}(\infty) + P_{2}P_{1}(\infty) + P_{2}P_{3}(\infty) )x^{2}\right.&\nonumber \\ 
&\left. - ( P_{1}P_{2}(\infty) + P_{3}P_{1}(\infty) + P_{3}P_{2}(\infty) )x - P_{3}P_{2}(\infty) \right]=0 \,.&
\end{eqnarray*}
Clearly, the whole equation contains the trivial solution $x=1$, i.e., $\eta^*=0$, while solving the third-order algebraic equation
\begin{eqnarray}\label{eq:receipe_2}
&P_{2}P_{3}(\infty)\,x^{3} + ( P_{1}P_{3}(\infty) + P_{2}P_{1}(\infty) + P_{2}P_{3}(\infty) )x^{2}&\nonumber \\
&- ( P_{1}P_{2}(\infty) + P_{3}P_{1}(\infty) + P_{3}P_{2}(\infty) )x = P_{3}P_{2}(\infty)&
\end{eqnarray}
provides us the other value of $\eta^{*} \neq 0$ that obeys the fluctuation relation $\mathcal{G}(j\eta^*)=1$. In this regard, by applying the well-known Routh-Hurwitz criterion to the polynomial (\ref{eq:receipe_2}), we can also prove that only one root of Eq.~(\ref{eq:receipe_2}) has positive real part different from $1$. In fact, according to the Routh-Hurwitz criterion, we recall that to each variation (permanence) of the sign of the coefficients of the first column of the Routh table corresponds to a root of the polynomial with a positive (negative) real part. In our case, there are always $2$ sign-permanences and only $1$ variation, for any possible value of the probabilities $P_{i}$ and $P_{f}(\infty)$. Being $\eta^*\propto -\ln x$, only the unique solution $x\neq 1$ with positive real part is physical, thus returning the unique non-trivial energy scale factor $\eta^*$ s.t.~$\mathcal{G}(j\eta^*)=1$.

\subsection{NV center with $\mathcal{H}\propto S_z$ subjected to dissipation}

As in the main text, for the case $\mathcal{H} \propto S_z$, we consider $\{ \ket{0},\ket{\pm 1} \}$ and $\{ E_0,E_{\pm 1} \}$ as the three eigenstates and eigenvalues of the qutrit.
Thus, assuming that $P_{\pm 1}(\infty)=0$ and $P_{0}(\infty) = 1$, one gets
\begin{align}
P_{0} + P_{1} e^{\eta^{*}E_{1}} + P_{-1} e^{\eta^{*}E_{-1}} &= 1\\ \Rightarrow \;
1-P_{1}-P_{-1} + P_{1} e^{\eta^{*}E_{1}} + P_{-1}e^{\eta^{*} E_{-1}} &= 1\\ \Rightarrow \; P_{1} (e^{\eta^{*} E_{1}}-1) + P_{-1} (e^{\eta^{*} E_{-1}}-1) &=0 \,.\label{eq:simplified_G}
\end{align}

\noindent
{\bf Low field, $\boldsymbol{\gamma_e B < \Delta}$:}\\
$E_{1}>0$ and $E_{-1}>0$ imply $\eta^{*}=0$ as the only possible solution.\\

As a proof, we can show that, by considering $\eta^{*} \ne 0$, Eq.~\eqref{eq:simplified_G} is not satisfied:
\begin{align}
\mathrm{If}\;\eta^{*}<0\;\;\Rightarrow & \;\;e^{\eta^{*} E_{1}}-1<0\; \mathrm{and}\; e^{\eta^{*} E_{-1}}-1<0 
\;\;\Rightarrow \;\; P_{1} (e^{\eta^{*} E_{1}}-1) + P_{-1} (e^{\eta^{*} E_{-1}}-1) < 0 \; \\
\mathrm{If}\;\eta^{*}>0\;\;\Rightarrow & \;\;e^{\eta^{*} E_{1}}-1>0\; \mathrm{and}\; e^{\eta^{*} E_{-1}}-1>0 
\;\;\Rightarrow \;\; P_{1} (e^{\eta^{*} E_{1}}-1) + P_{-1} (e^{\eta^{*} E_{-1}}-1) >0 \,.
\end{align}

\noindent
{\bf High field, $\boldsymbol{\gamma_e B > \Delta}$:}\\
Recalling that $E_{\pm 1} = \Delta \pm \gamma_e B$, in the limit of $\gamma_e B \gg \Delta$, Eq.~\eqref{eq:simplified_G} reduces to 
\begin{align}
P_{1} e^{\eta^{*} \gamma_e B} (1 - e^{-\eta^{*} \gamma_e B}) + P_{-1} (e^{-\eta^{*} \gamma_e B}-1) &=0\\ 
\Rightarrow \; (P_{1} e^{\eta^{*} \gamma_e B} - P_{-1}) (1 - e^{-\eta^{*} \gamma_e B}) &=0 \,.\label{eq:appendix_high_field}
\end{align}
The second part of the left-hand-side leads to the trivial solution $\eta^{*} = 0$. Instead, by putting together the conditions $(P_{1} e^{\eta^{*} \gamma_e B} - P_{-1}) = 0$ [from Eq.~(\ref{eq:appendix_high_field})] and $P_i = e^{-\beta E_i} / Z$, with $Z$ partition function (valid for any initial thermal state), we obtain:
\begin{align}
\frac{P_{-1}}{P_{1}} &=e^{\eta^{*} \gamma_e B}\\
\Rightarrow \eta^{*} &= \frac{1}{\gamma_e B} \ln {\left[e^{- \beta (E_{-1} - E_{+1})}\right]}\\
\Rightarrow \eta^{*} &= \frac{2 \gamma_e B \beta}{\gamma_e B} = 2 \beta \,.
\end{align}
This result is confirmed by numerical simulations to evaluate $\eta^{*}$, shown in Fig.~\ref{fig:Fig3} of the main text. Simulations are obtained by setting the initial inverse temperature $\beta$, and solving the condition $\mathcal{G} = 0$ for the characteristic function in the high field limit. In addition, for each selected initial temperature, we have iteratively derived $\eta^{*}$ while changing $B$ for $\gamma_e B \rightarrow \Delta_+$ by using each solution as initial guess for the subsequent simulation run.

\twocolumngrid
\bibliography{OQS-Biblio}

\end{document}